\documentclass[prd,twocolumn,showpacs,preprintnumbers,amsmath,amssymb,nofootinbib]{revtex4-1}

\usepackage{amsmath}
\usepackage{graphicx}
\usepackage{dcolumn}
\usepackage{xspace}
\usepackage{color}

\newcommand{\ep}{\epsilon}
\newcommand{\primed}{^{\prime}}
\def\beq#1\eeq{\begin{align}#1\end{align}}

\newcommand{\tev}{\,\mbox{TeV}}
\newcommand{\gev}{\,\mbox{GeV}}

\newcommand{\fig}[1]{Fig.~\ref{#1}}

\renewcommand{\Re}{\textrm{Re}\,}
\renewcommand{\Im}{\textrm{Im}\,}
\newcommand{\bra}[1]{ \langle {#1} | }
\newcommand{\ket}[1]{ | {#1} \rangle }
\newcommand{\lt}{\left}
\newcommand{\rt}{\right}
\newcommand{\eq}[1]{Eq.~(\ref{#1})}

\definecolor{BlueViolet}{rgb}{0.2, 0.00, 0.7}
\definecolor{Blue}{rgb}{0.15, 0.00, 0.9}
\usepackage[colorlinks=true,linkcolor=Blue,citecolor=Blue,urlcolor=BlueViolet,hyperfootnotes=false]{hyperref}

\begin{document}

\preprint{PSI-PR--17--03}
\preprint{TTP--17--013}
\title{\boldmath{$K\to \pi \nu\overline{\nu}$ in the MSSM in Light of the $\epsilon^{\prime}_K/\epsilon_K$ Anomaly}}

 \author{Andreas Crivellin}
\email{andreas.crivellin@cern.ch}
\affiliation{Paul Scherrer Institut, CH--5232 Villigen PSI, Switzerland}
\author{Giancarlo D'Ambrosio}
\email{gdambros@na.infn.it}
\affiliation{INFN-Sezione di Napoli, Via Cintia, 80126 Napoli, Italia}
\author{Teppei Kitahara} \email{teppei.kitahara@kit.edu} 
\affiliation{Institute for Theoretical Particle Physics (TTP), 
Karlsruhe Institute of Technology, Wolfgang-Gaede-Stra\ss e 1, 
 76128 Karlsruhe, Germany}
\affiliation{Institute for Nuclear Physics (IKP), Karlsruhe Institute of
  Technology, Hermann-von-Helmholtz-Platz 1, 76344
  Eggenstein-Leopoldshafen, Germany}
\author{Ulrich Nierste}
\email{Ulrich.Nierste@kit.edu}
\affiliation{Institute for Theoretical Particle Physics (TTP), 
Karlsruhe Institute of Technology, Wolfgang-Gaede-Stra\ss e 1, 
 76128 Karlsruhe, Germany}

\date{\today}

\begin{abstract}
  The Standard-Model (SM) prediction for the CP-violating {quantity}
  $\epsilon_K^\prime/\epsilon_K$ deviates from its measured value by 2.8 $\sigma$. It has been shown that this tension can be resolved within the Minimal Supersymmetric
  Standard Model (MSSM) through gluino-squark box diagrams, even if
  squarks and gluinos are much heavier than 1\tev. The rare decays
  $K_L \to \pi^0\nu\overline{\nu}$ and $K^+ \to \pi^+\nu\overline{\nu}$ are
  similarly sensitive to very high mass scales {and the first one also measures CP violation}.
In this article, we analyze {the} correlations between
$\epsilon^\prime_K/\epsilon_K$ and $\mathcal{B}(K_L \to
\pi^0\nu\overline{\nu})$ and $\mathcal{B}(K^+ \to
\pi^+\nu\overline{\nu})$ within the MSSM {aiming at an explanation of
$\epsilon_K^\prime/\epsilon_K$ via gluino-squark box diagrams}. The dominant MSSM contribution
to the $K \to \pi\nu\overline{\nu}$ branching fractions stems 
from box diagrams with squarks, sleptons, charginos, and neutralinos, and the pattern of the correlations is different from the widely studied $Z$-penguin scenarios. This is interesting in light of future {precision}
  measurements by KOTO and NA62 at J-PARC and CERN, respectively.
  We find $\mathcal{B}(K_L \to \pi^0\nu\overline{\nu})/\mathcal{B}^{\rm
    SM} (K_L \to \pi^0\nu\overline{\nu})\lesssim 2\,(1.2)$ and
  $\mathcal{B}(K^+ \to \pi^+\nu\overline{\nu})/\mathcal{B}^{\rm SM}(K^+
  \to \pi^+\nu\overline{\nu}) \lesssim 1.4\,(1.1)$, if all squark masses
  are above $1.5\,\tev$, gaugino masses obey GUT relations, and
if one allows for a fine-tuning at the $1\%\,(10\%)$ level for the
gluino mass.
 Larger values are possible for a tuned CP violating phase.
Furthermore, the sign of the MSSM contribution to
  $\epsilon_K^\prime$ imposes a strict
  correlation between $\mathcal{B}(K_L \to \pi^0\nu\overline{\nu})$ and
  the hierarchy between the masses $m_{\bar{U}}$,~$m_{\bar{D}}$ of the
  right-handed up-squark and down-squark: $\mbox{sgn}\, [\mathcal{B}(K_L
  \to \pi^0\nu\overline{\nu})-\mathcal{B}^{\rm SM} (K_L \to
  \pi^0\nu\overline{\nu}) ] = \mbox{sgn}\, (m_{\bar{U}}-m_{\bar{D}}) $. 
\end{abstract}

\maketitle

\renewcommand{\thefootnote}{\#\arabic{footnote}}
\setcounter{footnote}{0}
\section{Introduction\label{intro}}
Flavor-changing neutral current (FCNC) decays of $K$ mesons are
extremely sensitive to new physics (NP) and probe virtual effects of
particles with masses far above the reach of future colliders,
especially if the corresponding observable is CP violating. Prime examples of such observables are $\ep_K$ and
$\ep_K\primed$ measuring indirect and direct CP violation in $K\to
\pi\pi$ decays and also $K_L\to\pi^{0}\nu\overline{\nu}$. 
While indirect CP violation was already found in 1964 \cite{Christenson:1964fg}, 
it took 35 more years to establish a nonzero value of $\ep_K\primed$ in 1999 by the NA48 and KTeV 
Collaborations~\cite{epspexp}:
\begin{equation}
  \Re \lt. \frac{\epsilon_K\primed}{\epsilon_K}\rt|_{\rm exp} =  
    (16.6 \pm 2.3) \times
10^{-4}  \quad (\textrm{PDG~\cite{Olive:2016xmw}})
. \label{epsilon'exp}
\end{equation}
Until recently, large theoretical uncertainties precluded reliable
predictions for $ \Re(\epsilon _K'/\epsilon _K)$. Calculating {the}
hadronic matrix elements with the large-$N_c$ (dual QCD) method one
finds a Standard-Model (SM) value well below the experimental range given in \eq{epsilon'exp}
\cite{dual}. A major breakthrough has been the recent lattice-QCD
calculation of Ref.~\cite{Bai:2015nea}, which gives support to the
large-$N_c$ result. The current status is \cite{Kitahara:2016nld}
\begin{align}
\left. \frac{\epsilon_K\primed}{\epsilon_K}\right|_{\rm SM} = 
 (1.06 \pm 5.07) \times 10^{-4},
\label{eq:sm}
\end{align}
which is consistent with
  $(\epsilon_K\primed/\epsilon_K)_{\rm SM} =
 (1.9 \pm 4.5) \times 10^{-4}$ \cite{Buras:2015yba}.
Both results are based on the lattice numbers in
Refs.~\cite{Bai:2015nea, latticeA2} and further use CP-conserving $K\to
\pi\pi $ data to constrain some of the hadronic matrix elements
involved. The SM prediction in \eq{eq:sm} lies below the experimental
value in \eq{epsilon'exp} by 2.8\,$\sigma$.\footnote{Calculations using chiral  perturbation theory instead are consistent with both the measurement
  and \eq{eq:sm}, because they have larger errors~\cite{chipt}.}

This tension can be explained by NP effects like $Z^\prime$ gauge bosons~\cite{Buras:2015yca,Buras:2015kwd,Buras:2015jaq,Buras:2016dxz,Bobeth:2016llm}, models with modified $Z$-couplings~\cite{Buras:2015yca, Buras:2015jaq, Endo:2016tnu, Bobeth:2017xry}, by a right-handed coupling of quarks to the $W$~\cite{Cirigliano:2016yhc}, within the littlest Higgs model \cite{Blanke:2015wba},  but also within the Minimal Supersymmetric Standard Model  (MSSM)~\cite{Kitahara:2016otd, Endo:2016aws}. 

When pursuing such NP interpretations of the tension in
$\epsilon_K\primed$ it is natural to look for signatures in other $s\to
d$ transitions which are, in general, correlated in UV complete models. To
this end the rare decays $K_L\to \pi^0 \nu \overline{\nu}$  and  $K^+\to
\pi^+ \nu \overline{\nu}$  play an important role. Within the SM, the
branching ratios are predicted to 
be~\cite{Buras:2006gb,Brod:2010hi,Buras:2015qea}
\begin{align}
\mathcal{B}(K_L \to \pi^0 \nu \overline{\nu})_{\rm SM} &=& 
     (2.9\pm 0.2 \pm 0.0) \times 10^{-11}\,,\nonumber \\
 \mathcal{B}(K^+ \to \pi^+ \nu \overline{\nu})_{\rm SM} &=& 
    (8.3\pm 0.3 \pm 0.3) \times 10^{-11}. \label{eq:smk}
\end{align}
The first error summarizes the uncertainty from CKM parameters, the
second one denotes the remaining theoretical uncertainties (in
  $\mathcal{B}(K_L \to \pi^0 \nu \overline{\nu})_{\rm SM} $, $0.04$ is
  rounded to $0.0$).  The numbers in \eq{eq:smk} are based on the
best-fit result for the CKM parameters in {Ref.~}\cite{ckmfitter}.
Experimentally, we have \cite{Artamonov:2008qb}
\begin{equation}\label{EXP1}
\mathcal{B}(K^+ \to \pi^+ \nu \overline{\nu})_{\rm exp}=(17.3^{+11.5}_{-10.5})\times 10^{-11}\,,
\end{equation}
and the $90\%$ C.L. upper bound \cite{Ahn:2009gb},
\begin{equation}\label{EXP2}
\mathcal{B}(K_L \to \pi^0 \nu \overline{\nu})_{\rm exp}< 2.6\times 10^{-8}\,.
\end{equation}
In the future, these measurements will be significantly improved. The NA62 experiment at CERN \cite{Rinella:2014wfa,Romano:2014xda} is aiming to reach a precision of 10\,\% compared to the SM already in 2018. 
 In order to achieve $5\%$ accuracy more time is needed. 
Concerning $K_L \to \pi^0 \nu \overline{\nu}$, the KOTO experiment at J-PARC  aims in a first step at 
measuring  $\mathcal{B}(K_L \to \pi^0 \nu \overline{\nu})$ around the SM sensitivity~\cite{Komatsubara:2012pn,Shiomi:2014sfa}.
Furthermore, the KOTO-step2 experiment will aim at 100 events for the SM branching ratio, implying
a precision of 10\,\% of this measurement \cite{KOTO}.

\begin{figure*}[th]
\begin{center} 
\includegraphics[width=0.28\textwidth, bb = 0 0 1024 768]{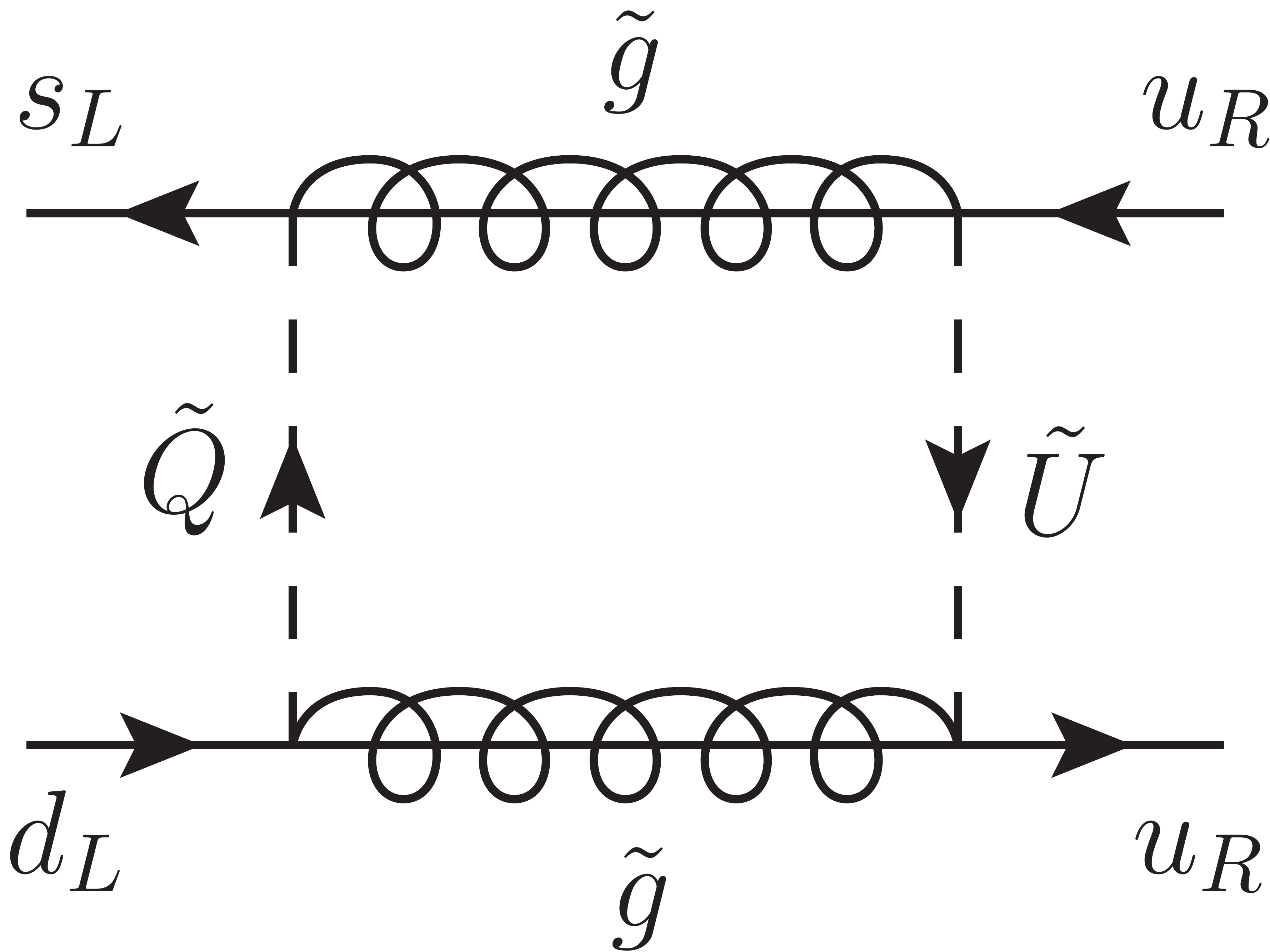}
\hspace{1cm}
\includegraphics[width=0.28\textwidth, bb = 0 0 1024 768]{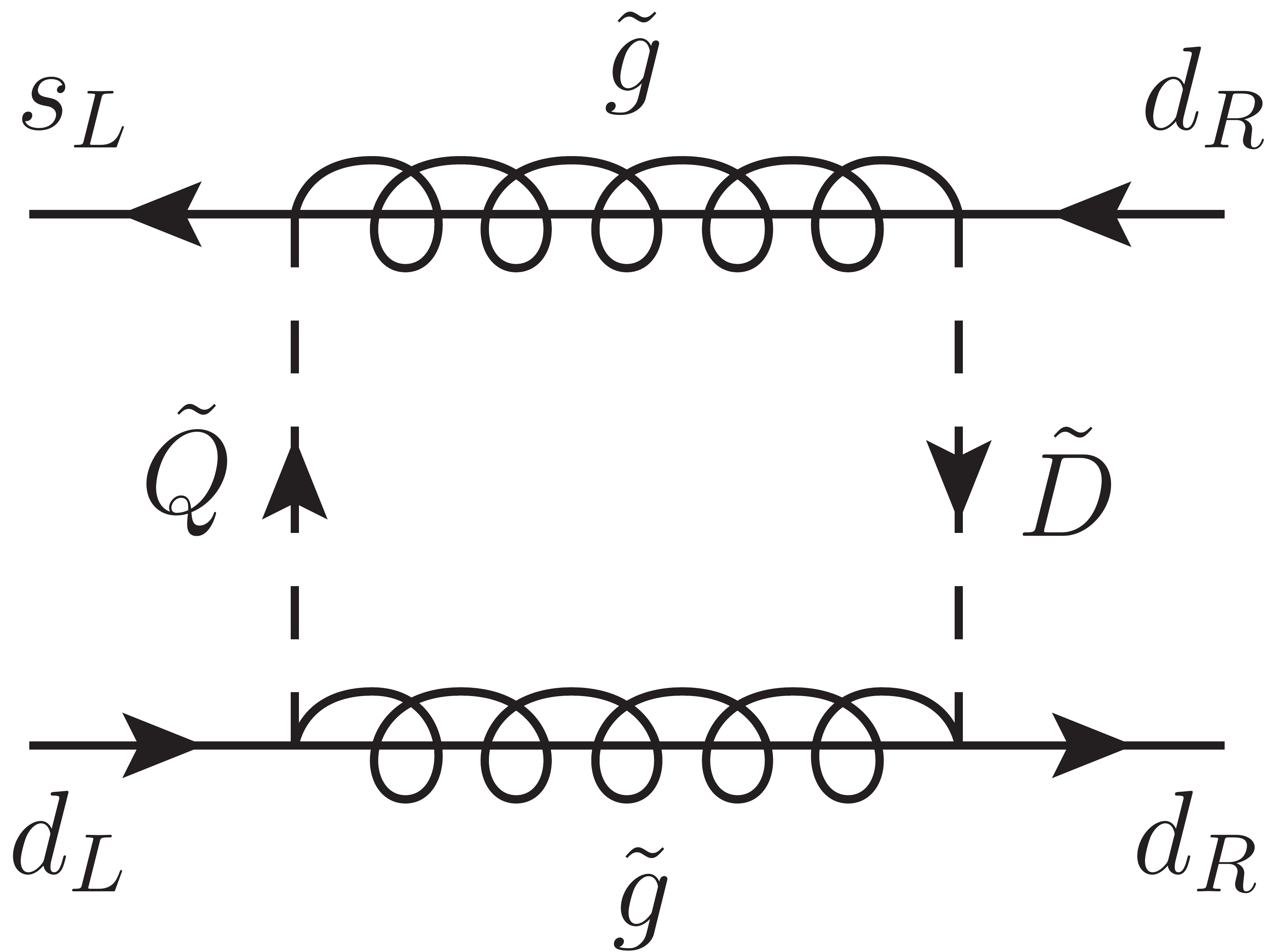}
\hspace{1cm}
\includegraphics[width=0.28\textwidth, bb = 0 0 1024 768]{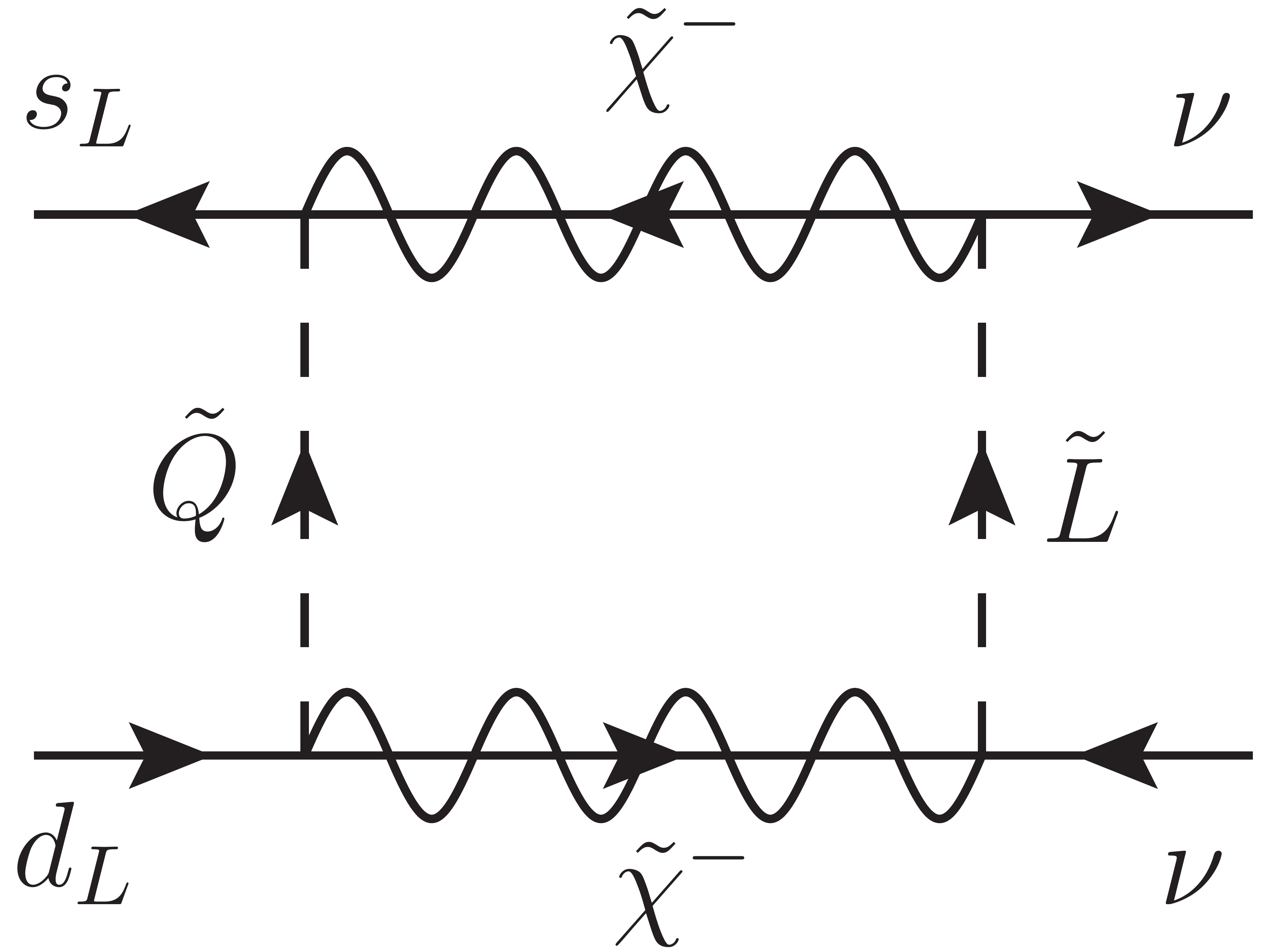}
\\ \vspace{0.5 cm}
\includegraphics[width=0.28\textwidth, bb = 0 0 1024 768]{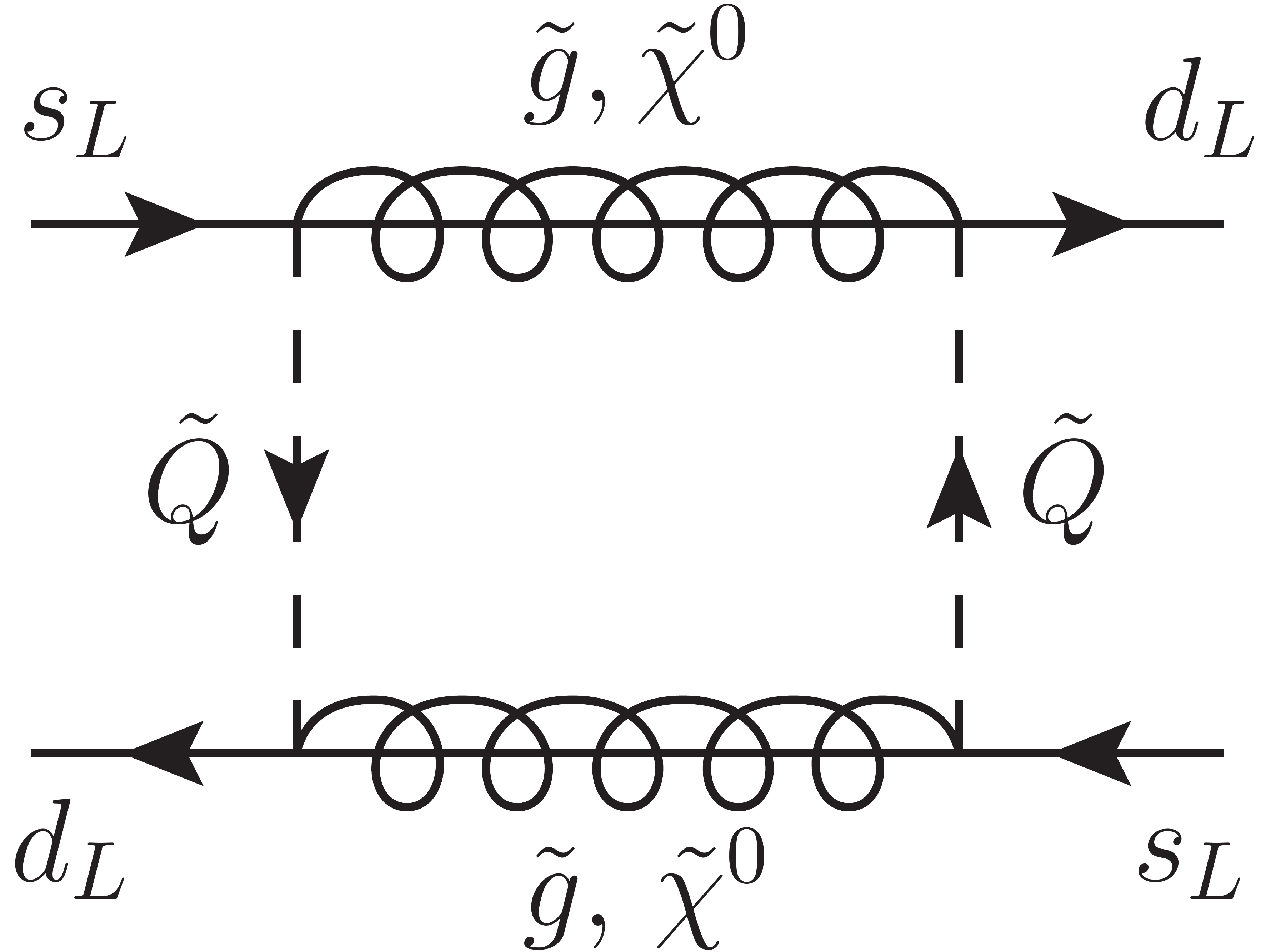}
\hspace{1cm}
\includegraphics[width=0.28\textwidth, bb = 0 0 1024 768]{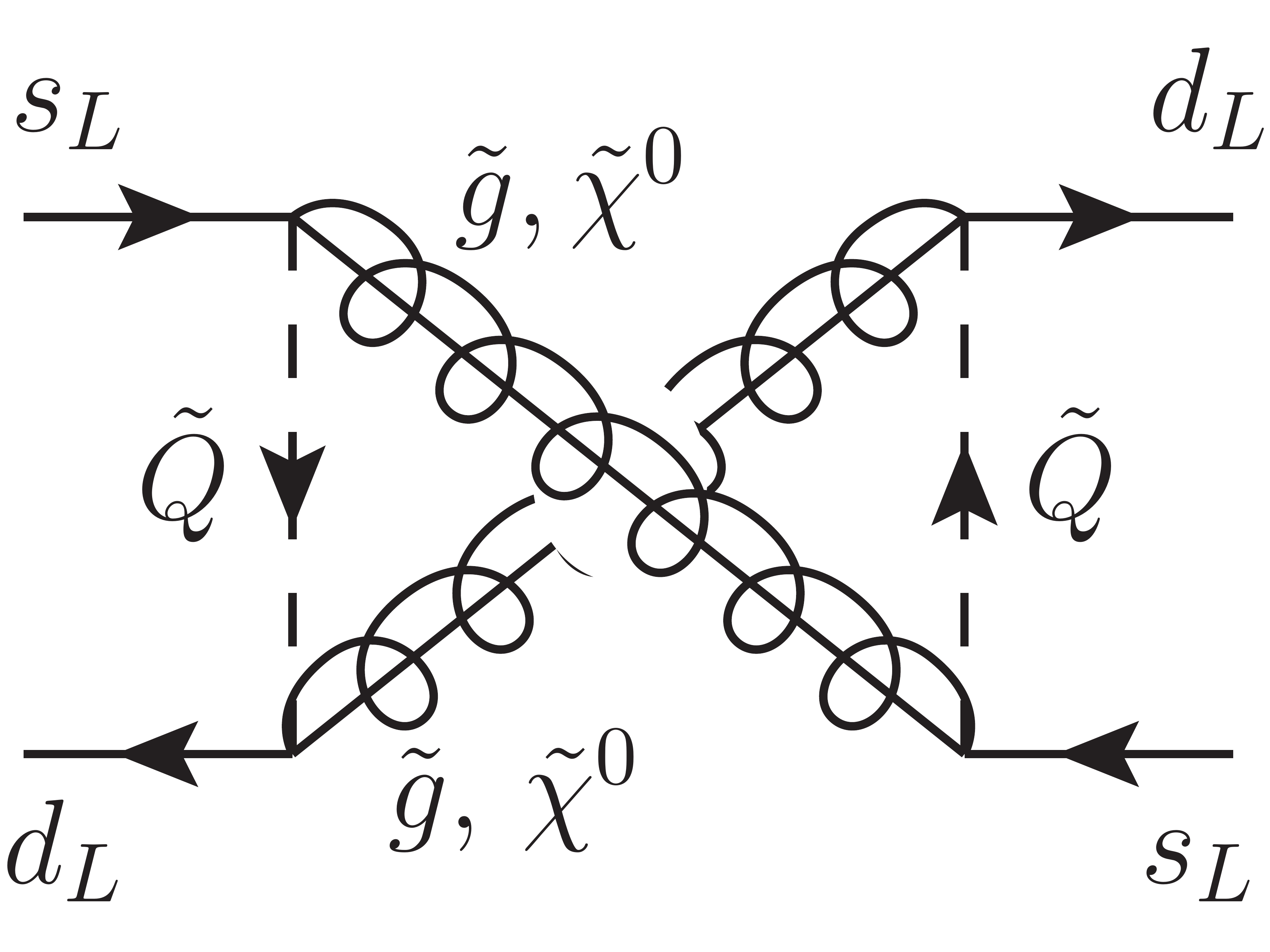}
\hspace{1cm}
\includegraphics[width=0.28\textwidth, bb = 0 0 1024 768]{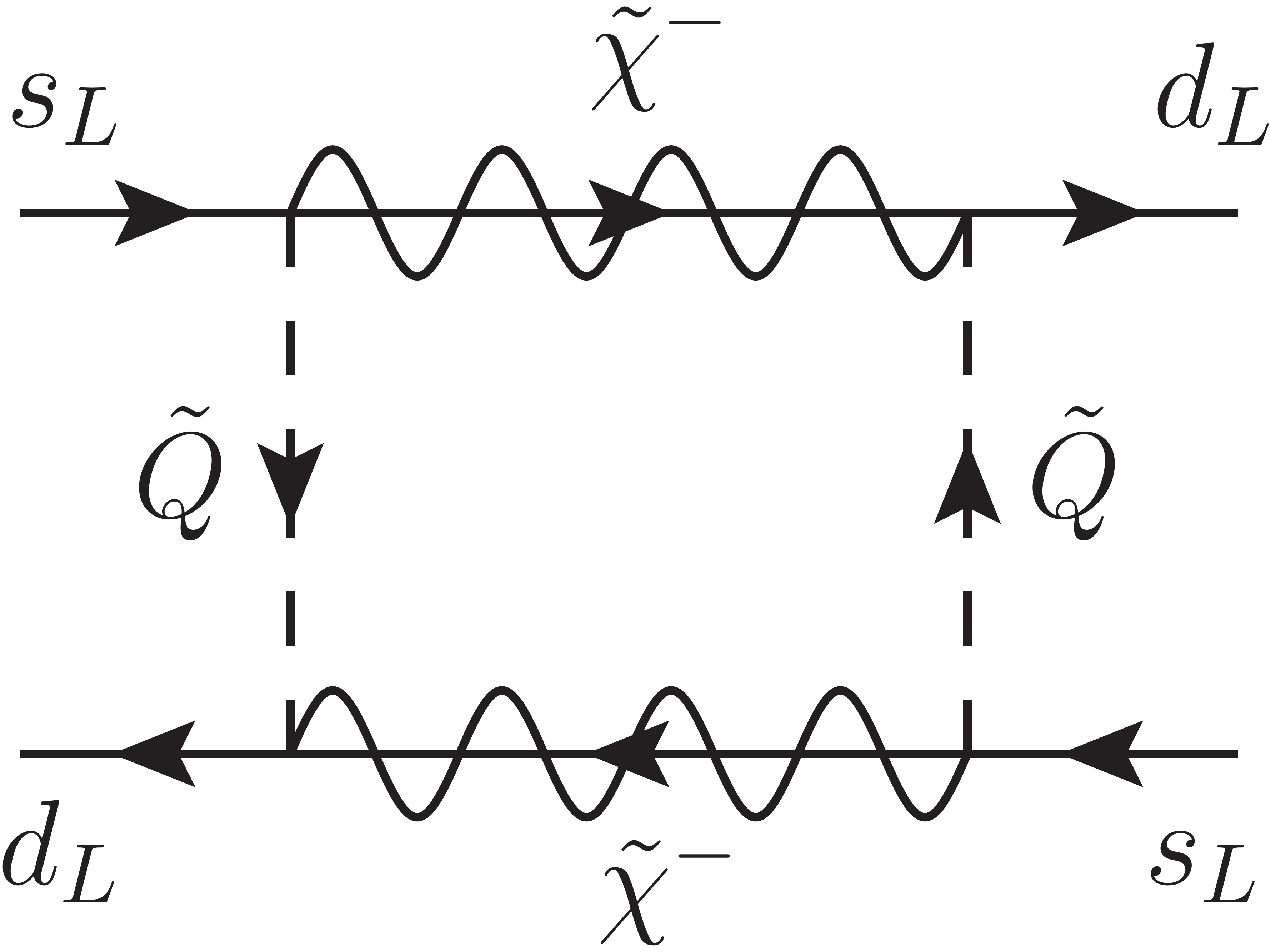}
\caption{Feynman diagrams of the dominant MSSM contributions to
    $\epsilon_K^\prime$, $K\to \pi \nu \overline{\nu}$, and $\epsilon_K$
    in our scenario. $\widetilde Q$ denotes a left-handed squark which
    is a down-strange mixture in our setup. $\widetilde{U}$ ($\widetilde{D}$)
    represents the right-handed up (down) squark.  $\widetilde g$,
    $\widetilde \chi^0$, and $\widetilde \chi^\pm$ stand for gluino,
    neutralino, and chargino, respectively, and $\widetilde L$ denotes a
    charged slepton.  {\em First row:} The first two box diagrams feed
    $\epsilon_K^\prime$ through $A_2$ in \eq{eq:mas} if $m_{U}\neq
    m_{D}$. The last diagram gives the ballpark of the MSSM contribution
    to $\mathcal{B}(K\to \pi\nu\overline{\nu})$.  {\em Second row:} MSSM
    contributions to $\epsilon_K$.
\label{fig:diagrams}}
\end{center}
\end{figure*}

In our MSSM scenario---in which {the} desired effect in
$\epsilon_K\primed$ is generated via gluino-squark boxes
\cite{Kitahara:2016otd}--- correlations with $\mathcal{B}(K_L\to \pi^0
\nu \overline{\nu})$ and $\mathcal{B}(K^+\to \pi^+ \nu \overline{\nu})$
are not unexpected, since sizable box contributions also occur in these
rare decays \cite{Buras:2004qb} (see \fig{fig:diagrams}).
Reference~\cite{Kitahara:2016otd} {achieves} sizable effects in
$\epsilon_K\primed$~\cite{gkn} together with a simultaneous efficient
suppression of the supersymmetric QCD contributions to $\epsilon_K$
\cite{Crivellin:2010ys}. The suppression occurs because crossed and
uncrossed gluino box diagrams cancel {if} the gluino mass {is} roughly
1.5 times the squark masses. With appropriately large left-left squark
mixing angle and {a} CP phase one can reconcile the measurements of
$\epsilon_K$ and $\Delta M_K$ with the large value in \eq{epsilon'exp}
and squark and gluino masses in the multi-TeV range, so that there is no
conflict with collider searches.

However, there is no such cancellation in the (dominant) chargino box
contribution to $K_L\to \pi^0 \nu \overline{\nu}$ and $K^+\to \pi^+ \nu
\overline{\nu}$ which permits potentially large effects.

This article is organized as follows: In the next section, we will review
$\epsilon^\prime_K/\epsilon_K$ and $K\to \pi \nu \overline{\nu}$ within
the MSSM. In Sec.~\ref{pheno}, we then perform the phenomenological
analysis highlighting the correlations before we conclude in
Sec.~\ref{conclusions}.

\section{Preliminaries}
$\epsilon_K^\prime/\epsilon_K$ is given by~\cite{Buras:2015yba}
\beq
\frac{\epsilon_K^\prime}{\epsilon_K} = \frac{\omega_{+}}{\sqrt{2}
  {|}\epsilon_K^{\textrm{exp}}{|} \Re A_0^{\textrm{exp}} }
\left\{ \frac{\textrm{Im} A_2 }{\omega_{+}} - \left( 1-
    \hat{\Omega}_{\textrm{eff}} \right) \textrm{Im} A_0 \right\},
\label{eq:mas}
\eeq%
with $\omega_{+}= (4.53 \pm 0.02)\times10^{-2}$, $|\epsilon_K^{\rm exp}|=(2.228\pm 0.011)\cdot 10^{-3}$,
$\hat{\Omega}_{\textrm{eff}} = (14.8\pm 8.0)\times 10^{-2}$, and the
amplitudes $A_I = \langle (\pi \pi)_I | \mathcal{H}^{\left|\Delta
    S\right| = 1} | K^0 \rangle$ involving the effective $|\Delta S|=1$
Hamiltonian $ \mathcal{H}^{\left|\Delta S\right|=1}$. 
Short-distance physics enters $\Im A_0 $ and $\Im A_2 $ through the
Wilson coefficients in $ \mathcal{H}^{\left|\Delta S\right|=1}$. The SM
prediction of the renormalization-group (RG) improved Wilson
coefficients is known to the next-to-leading order (NLO) of QCD and QED corrections 
\cite{epspnlo} and the next-to-next-to-leading-order QCD calculation  is
underway \cite{Cerda-Sevilla:2016yzo}. Equation~(\ref{eq:sm}) 
is based on a novel analytic formula for the NLO RG evolution.  

The Wilson coefficients multiply the four-quark operators $Q_j$ whose
hadronic matrix elements $\bra{(\pi\pi)_I} Q_j \ket{K^0}$ must be
calculated by nonperturbative methods. For some time these calculations
for the matrix elements entering $\Im A_2$ are in good shape, thanks to
precise results from lattice QCD \cite{latticeA2}. However, $\Im A_0$
has become tractable with lattice QCD only recently \cite{Bai:2015nea}.

CP-conserving data determine $\Re A_0$ and $\omega_+$ in
\eq{eq:mas}. $\omega_+$ is essentially equal to the ratio $\Re A_2/\Re
A_0$, except that it is calculated from charged rather than neutral kaon
decays. The smallness of $\omega_+$ encodes the famous ``$\Delta
I=1/2$'' rule $\Re A_0\gg \Re A_2$. It leverages the $\Im A_2$ term in
\eq{eq:mas} and leads to the above-mentioned high sensitivity of
$\ep_K\primed$ to new physics in this amplitude.

Following the approach of Ref.~\cite{Kitahara:2016otd} we aim at
explaining the discrepancy in $\epsilon_K^\prime/\epsilon_K$ with
contributions to the Wilson coefficients $c_{1,2}^{\prime
  q}$. Therefore, we need the flavor (and CP) violation in the
left-handed squark sector while the mass difference between the
right-handed up- and down-squarks accounts for the necessary isospin
violation.

The small errors in \eq{eq:smk} show that the $K\to\pi\nu\overline{\nu}$ branching ratios are theoretically very
clean.  While $K_L\to\pi^0\nu\overline{\nu}$ is only sensitive to
the CP violating part of the amplitude, $K^+\to\pi^+\nu\overline{\nu}$
is dominated by the CP conserving part. In principle, many diagrams contribute
to $K\to\pi\nu\overline{\nu}$ in the MSSM with generic sources of flavor
violation~\cite{Buras:2004qb}. However, since we are interested in a
scenario with $s-d$ flavor violation in the left-handed squark sector,
chargino-box contributions are numerically most important.

\begin{figure*}[th]
\begin{center} 
\includegraphics[width=0.46\textwidth, bb= 0 0 360 362]{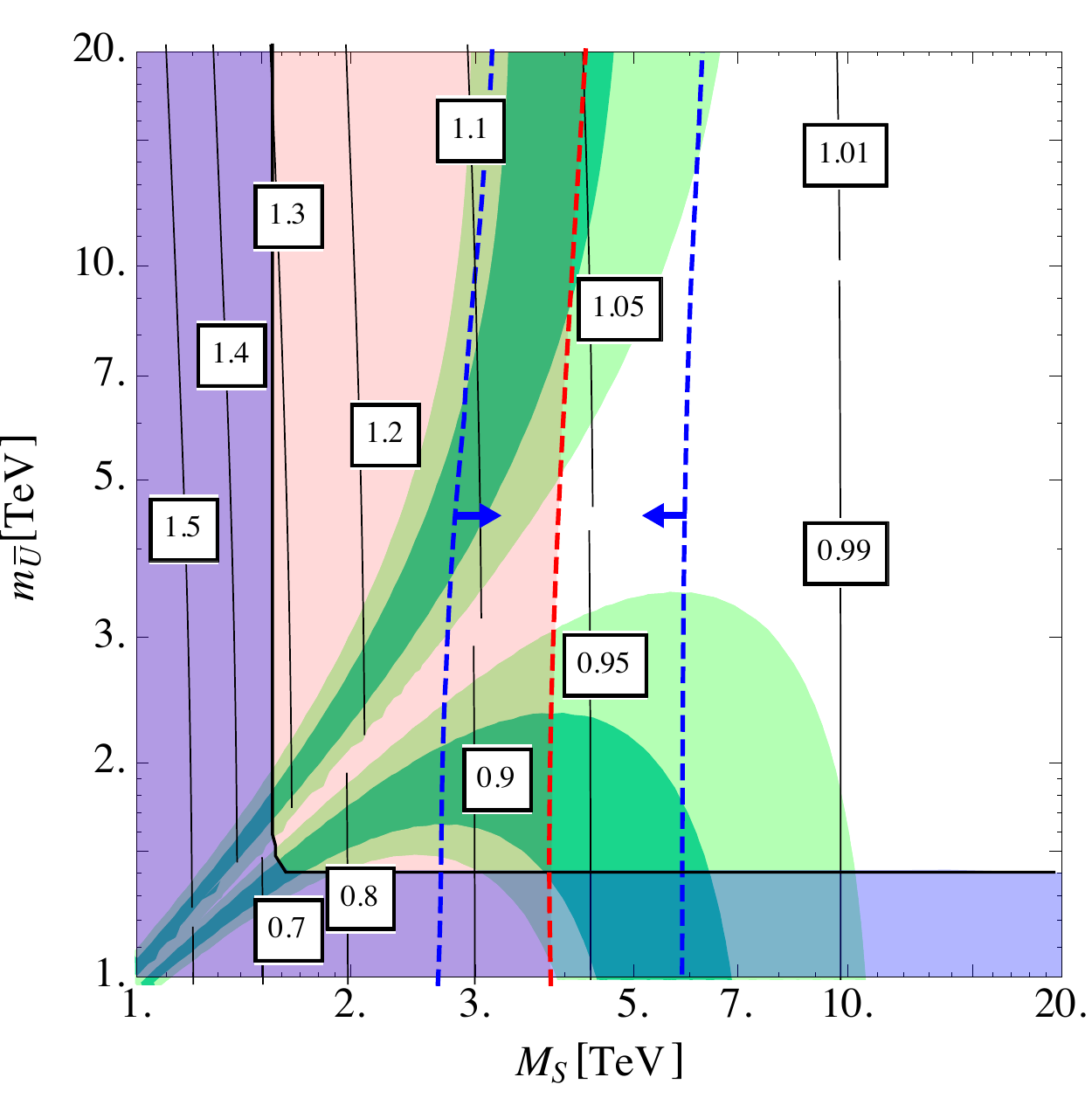}
\includegraphics[width=0.45\textwidth, bb= 0 0 354 352]{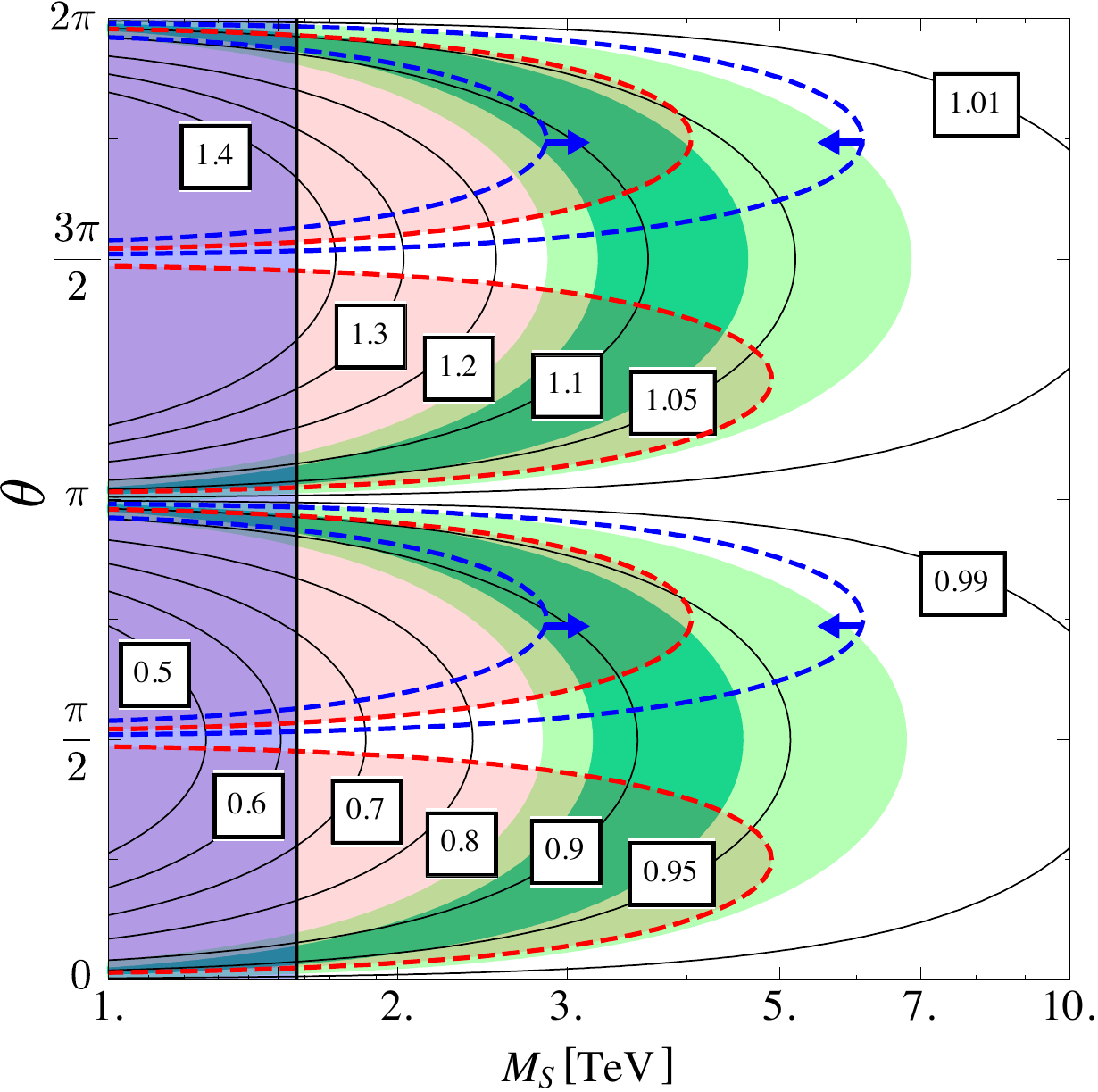}
\caption{Contours of $\mathcal{B}(K_L\to\pi^{0}
  \nu\overline{\nu})/\mathcal{B}^{\rm SM} (K_L \to
  \pi^0\nu\overline{\nu})$. The $\epsilon_K^\prime/\epsilon_K$
  discrepancy is resolved at the $1\,\sigma$\,($2\,\sigma$) level within
  the dark (light) green region.  The red shaded region is excluded by
  $\epsilon_K$ at 95\,\% C.L. using the inclusive value $|V_{cb}|$ ,
  while the region between the blue-dashed lines can explain the
  $\epsilon_K$ discrepancy which is present if the exclusive
  determination of $V_{cb}$ is used \cite{Bailey:2015tba}.  
The blue shaded region is excluded by the current LHC results {from CMS and ATLAS}~\cite{ATLAS:2016kts,CMS:2016xva,Aad:2015iea}.
  $M_3 / M_S =
  1.5$, $m_L =300\,\gev$ and GUT relations among gaugino masses are
  used. In the \emph{left} plot, $\Delta_{Q,12} = 0.1 \exp(- i \pi
  /4)$ for $m_{\bar{U}} > m_{\bar{D}} =m_Q= M_S $ (upper branch) and
$\Delta_{Q,12} = 0.1 \exp( i 3 \pi /4)$ for $m_{\bar{U}}
  <  m_{\bar{D}} = m_Q= M_S $ (lower branch). 
In the \emph{right} plot, $| \Delta_{Q,12}| = 0.1$ is used, $m_{\bar{D}}=2 m_{\bar{U}} = 2 m_Q = 2 M_S$ (for $0 < \theta < \pi)$ and $m_{\bar{U}}=2 m_{\bar{D}} = 2 m_Q = 2 M_S$ (for $\pi < \theta <
  2 \pi)$.}
\label{fig:contour}
\end{center}
\end{figure*}

\section{Phenomenological analysis}
\label{pheno}

Although the correlations between $\epsilon_K^\prime/\epsilon_K$ and $K
\to \pi \nu \overline{\nu}$ in the MSSM have already been discussed in
detail in
Refs.~\cite{Buras:1998ed,Buras:2004qb,Tanimoto:2016yfy,Endo:2016aws},
our study has several novelties. First of all,
Refs.~\cite{Buras:1998ed,Buras:2004qb} were written before the
appearance of the $\epsilon_K^\prime$ anomaly, while we take into
account the implication of the current deviation from the SM
prediction. With the progress on the SM prediction, $\epsilon_K^\prime$
implies a much sharper constraint on the MSSM parameters, {resulting}
in tighter bounds on the deviations of $\mathcal{B}(K \to \pi \nu
\overline{\nu})$ from the SM prediction. In addition, in our analysis we
employ $m_{\bar{U}}\neq m_{\bar{D}}$ to generate large gluino box ({\it
  Trojan} penguin) \cite{gkn} contributions to $\epsilon_K^\prime$,
while Refs.~\cite{Tanimoto:2016yfy,Endo:2016aws} enhance
$\epsilon_K^\prime$ through $Z$ penguins. Furthermore, we consider the
latest LHC limits on the supersymmetric (SUSY)
masses~\cite{TheATLAScollaboration:2013hha, Aad:2015iea,
  ATLAS:2016kts,CMS:2016xva}.

Defining the bilinear terms for the squarks as $M^2_{X,ij} = m_X^2
(\delta_{ij} + \Delta_{X,ij})$ for $X=Q,\bar U, \bar D$, the
numerically relevant parameters entering $\epsilon_K^\prime$,
$\epsilon_K$ and $K \to \pi \nu \overline{\nu}$ in our analysis are \beq
m_Q,~|\Delta_{Q,12}|,~\theta,~M_3,~M_2,~M_1,~m_{\bar{U}}/m_{\bar{D}},~m_L.
\eeq Here $m_Q$ is the universal mass parameter for the bilinear terms
of the left-handed squarks which we define in the down-quark basis
(i.e. the up-squark mass matrix is obtained via a CKM rotation from
$M^2_{Q}$). $\theta \equiv \textrm{arg}(\Delta_{Q,12})$, $M_3$ is the
gluino mass, $M_2$ ($M_1$) the wino (bino) mass, and $m_L$ is the
(universal) mass for the left-handed sleptons, respectively. The
trilinear $A$-terms as well as the off-diagonal elements of the
bilinear terms $\Delta_{X,ij}$ are set to $0$ except for $\Delta_{Q,12}$
which generates the required flavor and CP violation in our setup. The
values of the other (SUSY) parameters barely affect our
results.\footnote{We use the fixed values $\tan \beta =10$, $\mu = M_A =
  m_Q$, $A_{ij} =0$. We also fix $B_G = 1$, which parameterizes the
  matrix element of the chromomagnetic penguin operator $Q_{8g}$.}

The SUSY contribution to $\epsilon_K$ ($\epsilon_K^{\rm SUSY}$) and
$\Delta M_K$, {originates from} one-loop boxes with all possible
combinations of gluinos, winos, and binos.  For $K^+\to\pi^{+}
\nu\overline{\nu}$ and $K_L \to \pi^0\nu\overline{\nu}$ we take into
account all MSSM one-loop contributions~\cite{Buras:2004qb}. However,
numerically the chargino boxes turn out to be by far dominant in our
setup.  In $\epsilon'_K /\epsilon_K$, we include all SUSY QCD (SQCD)
contributions as well as $Z$-penguin contributions originating from
chargino diagrams to the $I=0,\,2$ amplitudes with hadronic matrix
elements evaluated at $1.3\,\gev$
\cite{Kitahara:2016nld,Kitahara:2016otd}.  {In the calculation of all
  contributions, we perform an exact diagonalization of the squark mass
  matrices.}

\begin{figure*}[t]
\begin{center} 
\includegraphics[width=0.44\textwidth, bb= 0 0 360 381]{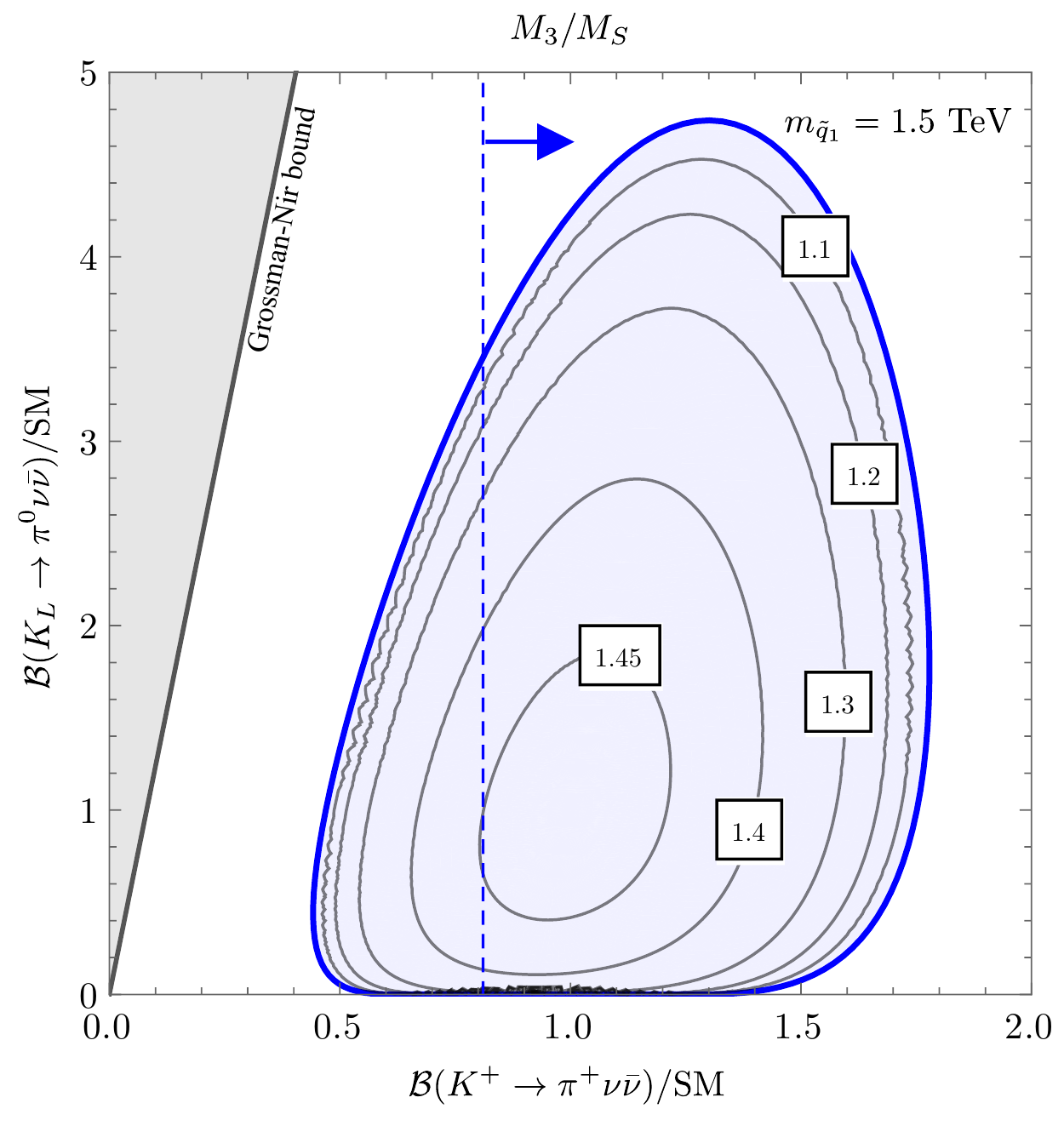}
\includegraphics[width=0.46\textwidth, bb= 0 0 360 365]{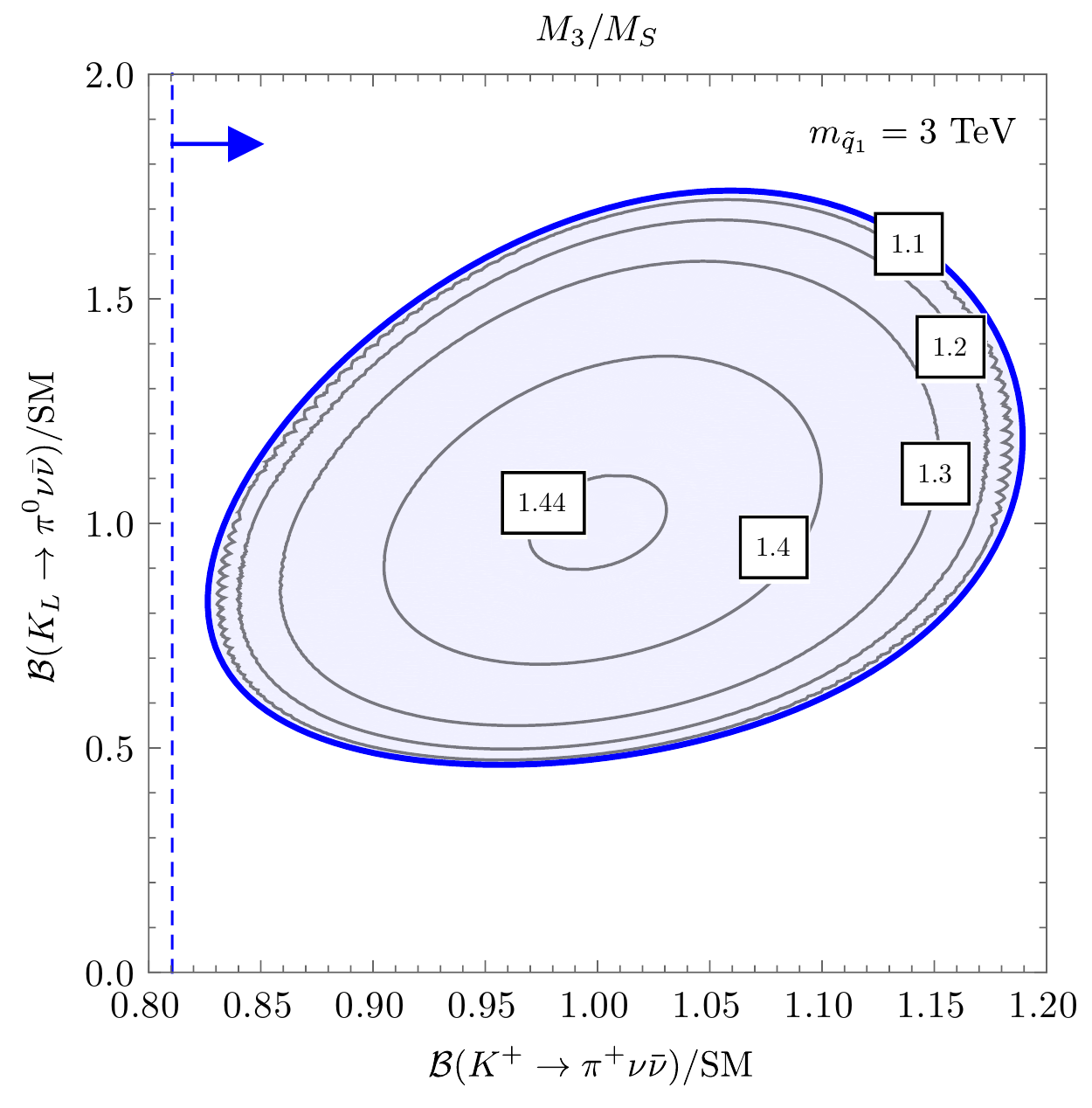}
\caption{Allowed region in the $\mathcal{B}(K_L\to\pi^{0}
  \nu\overline{\nu})$--$\mathcal{B}(K^{+}\to\pi^{+} \nu\overline{\nu})$
  plane.  ``SM" in the axis labels represents the corresponding value of
  the branching ratio within the SM.  The contours show the values of
  $M_3 / M_S$ which is needed to cancel the SUSY contributions to
  $\epsilon_K$. In the \emph{left} (\emph{right}) panel, the lightest
  squark mass is fixed at 1.5\,(3)\,\tev.  The gray shaded region is the
  Grossman-Nir bound~\cite{Grossman:1997sk}. The right sides of the blue
  dashed lines are the experimental result for
  $\mathcal{B}(K^{+}\to\pi^{+} \nu\overline{\nu})$ { given in \eq{EXP1}}.  }
\label{fig:M3MScontour}
\end{center}
\end{figure*}

In the SM contributions, we fix the relevant CKM elements to their
  best-fit values \cite{ckmfitter}, in particular we set
  $V_{td}^*V_{ts}= (- 3.22 + 1.41 i)\cdot 10^{-4}$. In this way, we
  assume that the MSSM contributions to the standard unitarity-triangle
  analysis are small, so that the change in $V_{td}^*V_{ts}$ is
  unimportant compared to the explicit MSSM contributions to
  $\epsilon_K^\prime$ and $\mathcal{B}(K\to \pi \nu \overline{\nu})$. This
  is justified  in typical MSSM scenarios with generic flavor violation.

  First, we show a typical prediction for $\mathcal{B}(K_L\to\pi^{0}
  \nu\overline{\nu})/\mathcal{B}^{\rm SM} (K_L \to
  \pi^0\nu\overline{\nu})$ as a function of the squark masses in
  the left panel of Fig.~\ref{fig:contour}. Here we assume universal
  diagonal elements for the left-handed and right-handed down squark mass
  matrices $M_S = m_Q = m_{\bar{D}}$ and use $m_L =300\,\gev$. We also
  choose $\Delta_{Q,12} = 0.1 \exp(- i \pi /4)$~($0.1 \exp( i 3 \pi /4)$)
  for $m_{\bar{U}} > M_S $~($m_{\bar{U}} < M_S $) regions to obtain a
  positive contribution to $\epsilon^{\prime}_K$.  We impose $M_3/M_S =
  1.5$ in order to obtain an efficient suppression of $\epsilon_K^{\rm
    SUSY}$~\cite{Crivellin:2010ys,Kitahara:2016otd}. In addition, the GUT
  relation for $M_2$ and $M_1$ are imposed.  The
  $\epsilon_K^\prime/\epsilon_K$ discrepancy between \eq{epsilon'exp}
  and the second prediction in \eq{eq:sm} is resolved at
  $1\,\sigma$\,($2\,\sigma$) within the dark (light) green region.  The
  red shaded region is excluded by $\epsilon_K$ at 95\,\% C.L. if the
  inclusive value of $|V_{cb}|$ is used, while the region between the
  blue-dashed lines can explain the $\epsilon_K$ discrepancy present if
  the exclusive determination of $V_{cb}$ is used
  \cite{Bailey:2015tba}.\footnote{The difference compared to Fig.\,4 of
    Ref.\,\cite{Kitahara:2016otd} comes from $\Delta_{Q,13,23}$.}  Note
  that $\theta = \pm \pi /4$ maximizes the effect in $\epsilon_K^{\rm
    SUSY}$, while the SUSY contributions to $\epsilon^{\prime}_K /
  \epsilon_K $ is maximized at $\theta = \pm \pi /2$ resulting instead
  in a vanishing effect in $\epsilon_K^{\rm SUSY}$.  The blue shaded
  region is excluded by the current LHC
  results~\cite{ATLAS:2016kts,CMS:2016xva,Aad:2015iea}.  {Here, in order
    to be conservative, we use the most stringent one, i.e. we
    maximize the bound which is a function of the neutralino mass.}  In
  this setup, we find that $\mathcal{B}(K_L\to\pi^{0}
  \nu\overline{\nu})/\mathcal{B}^{\rm SM} (K_L \to
  \pi^0\nu\overline{\nu}) \simeq 1.05$--$1.1$ is predicted in light of
  the $\epsilon_K^\prime/\epsilon_K$ discrepancy (and the potential
  $\epsilon_K$ discrepancy) if $m_{\bar{U}} > m_{\bar{D}}$.

In the right panel of Fig.~\ref{fig:contour}, the {dependence on the CP-violating phase ($\theta$)} is shown. Here, we chose $| \Delta_{Q,12}| = 0.1$, and $m_{\bar{D}}=2 m_{\bar{U}} = 2 m_Q = 2 M_S$ ($m_{\bar{U}}=2 m_{\bar{D}} = 2 m_Q = 2 M_S$ ) for $0 < \theta < \pi  $ ($\pi < \theta <
  2 \pi)$.  It {can be seen that if $\theta$ is close $\pm \pi/2 $, the constraint from $\epsilon_K$ is weakened while $\epsilon^{\prime}_K $ as well as $\mathcal{B}(K_L\to\pi^{0}
\nu\overline{\nu})$ is enhanced.}

Next, let us investigate upper and lower limits on
$\mathcal{B}(K_L\to\pi^{0} \nu\overline{\nu})$ and
$\mathcal{B}(K^+\to\pi^{+} \nu\overline{\nu})$.  In the following
analysis, we fix the slepton mass close to the experimental limit ($m_L
= 300\,\gev$)~\cite{TheATLAScollaboration:2013hha} and use GUT
relations among all three gaugino masses. Therefore, when one fixes the
  lightest squark mass, the relevant free parameters are only \beq
|\Delta_{Q,12}|,~\theta,~M_3,~m_{\bar{U}}/m_{\bar{D}}, \eeq with $0 <
|\Delta_{Q,12}| < 1$ and $0 < \theta < 2 \pi$.
In Fig.~\ref{fig:M3MScontour}, the blue solid line encloses the
maximally allowed region in the $\mathcal{B}(K_L\to\pi^{0}
\nu\overline{\nu})$--$\mathcal{B}(K^{+}\to\pi^{+} \nu\overline{\nu})$
plane (normalized by their SM values).  The maximal values are obtained
whenever the SUSY contributions to the $\Delta S = 2$ amplitude exactly
cancel. The contour lines in the figures show the required value of
$M_3/M_S$ (imposing again GUT relations) for this cancellation. The
maximal and minimal values for $\mathcal{B}(K\to\pi \nu\overline{\nu})$
are obtained by the decoupling one of the left-handed mixed down-strange
squark while simultaneously maximizing their mixing. Since we assume
equal diagonal entries of the bilinear terms this corresponds to the
limit $m_Q \to \infty$ and $|\Delta_{Q,12}| \to 1$ which implies one
light squark which is an equal admixture of the first and second
generation of interaction eigenstates. Note that these results are
independent of $m_{\bar{U}}/m_{\bar{D}}$, but $m_{\bar{U}}/m_{\bar{D}}$
is important when considering the correlation with
$\epsilon^{\prime}_K/\epsilon_K$.  In the left and right panels, the
lightest squark mass is fixed to 1.5\,\tev~and 3\,\tev, respectively.
The latest searches for first-generation squarks at the LHC {imply}
$m_{\tilde{q}_1} \gtrsim 1.4\,\tev$ if the gluino is heavy and the
neutralino is light \cite{ATLAS:2016kts, CMS:2016xva}.  We find that the
upper allowed values for the branching ratios differ significantly from
the SM predictions. However, in order to achieve these maximal values,
severe fine-tuning of the gluino mass (with respect to the squark
masses) or tuning of the CP violating phase is necessary:
e.g. around $\theta = 3 \pi/2$, $\epsilon_K^{\rm SUSY}$ is much
suppressed while $\mathcal{B}(K_L\to\pi^{0} \nu\overline{\nu})$ is
enhanced.

\begin{figure*}[t]
\begin{center}
\includegraphics[width=0.45\textwidth, bb= 0 0 356 350]{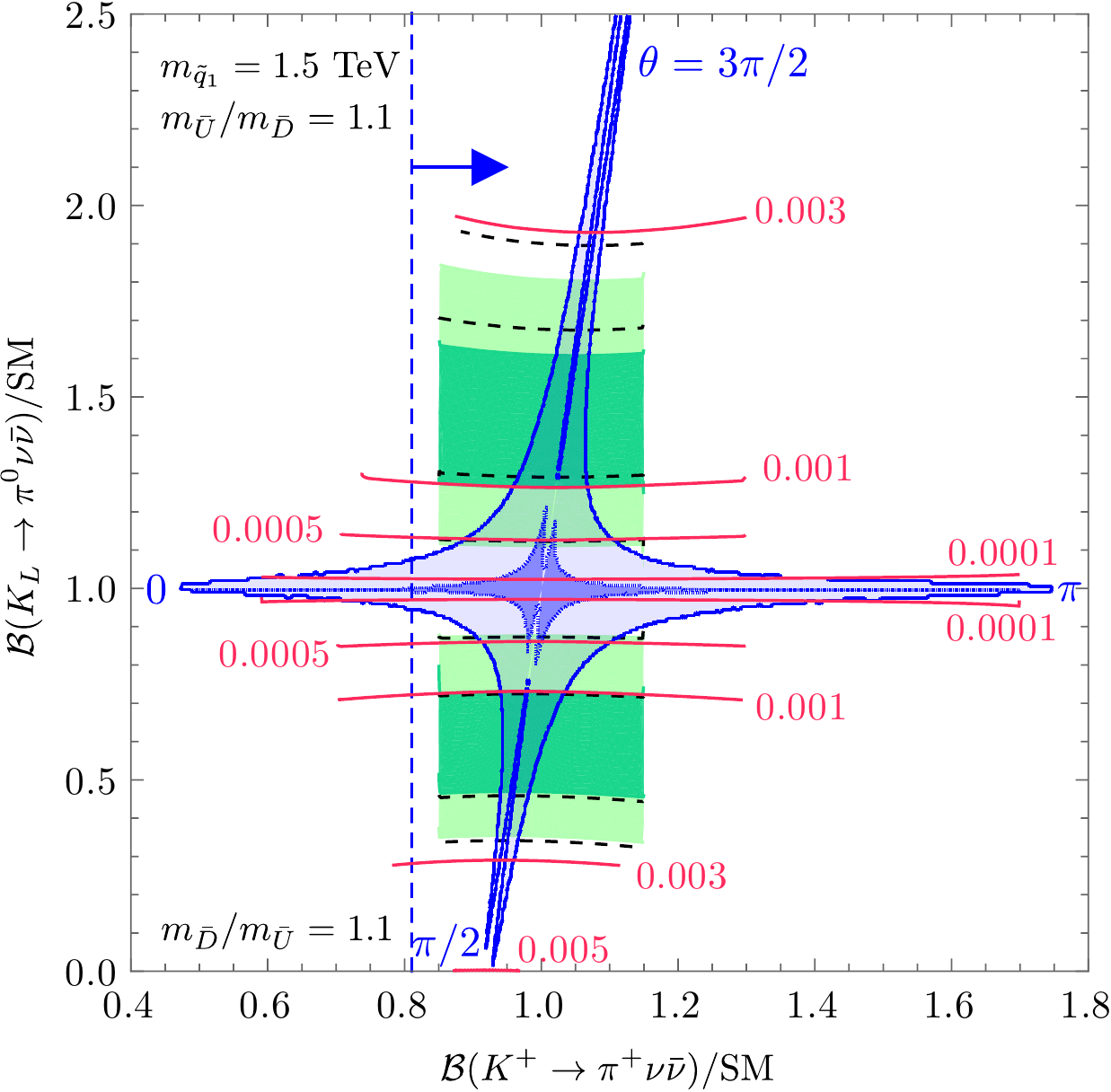}
\includegraphics[width=0.45\textwidth, bb= 0 0 352 345]{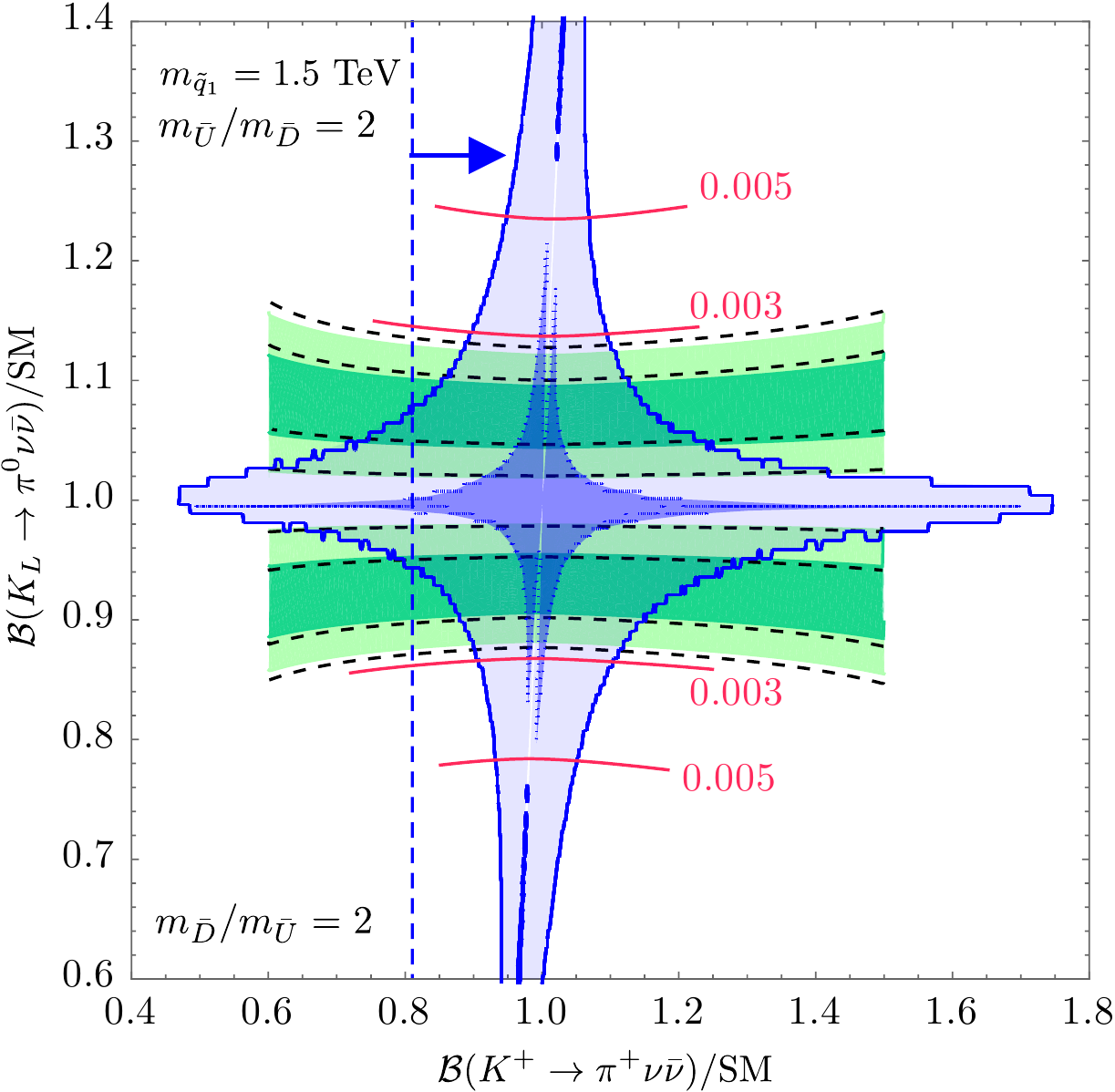}
\caption{ The light (dark) blue region requires a milder parameter
  fine-tuning than 1\,\%\,(10\,\%) of the gluino mass compared to
  the value of Fig.~\ref{fig:M3MScontour} and a milder parameter tuning
  than 1\,\%\,(10\,\%) of the deviation of the CP violating phase from
$\pm \pi/2$. The plot suggests that by tuning the CP violating phase
  $\theta$ close to $\pm \pi/2$ one can relax the fine-tuning of the
  gluino mass and still amplify the branching ratios. 
The red contour represents the SUSY contributions to
$\epsilon^{\prime}_K / \epsilon_K$, and the
$\epsilon_K^\prime/\epsilon_K$ discrepancy is resolved at
$1\,\sigma$\,($2\,\sigma$) within the dark (light) green region. The
black dashed lines show the projected shifts of the boundaries of the
green regions when the gluino is assumed to be 10\,\% heavier.  The
lightest squark mass is fixed to 1.5\,\tev.  In the \emph{left} panel,
$m_{\bar{D}}/m_{\bar{U}} = 1.1$ ($m_{\bar{U}}/m_{\bar{D}} = 1.1$) is
used for $0 < \theta < \pi$ ($\pi < \theta < 2 \pi$) to obtain a
positive SUSY contribution to $\epsilon^{\prime}_K / \epsilon_K$. While,
$m_{\bar{D}}/m_{\bar{U}} = 2$ ($m_{\bar{U}}/m_{\bar{D}} = 2$) is used
for $0 < \theta < \pi$ ($\pi < \theta < 2 \pi$) in the \emph{right}
panel.  The region on the right side of the blue dashed lines are
allowed by the current experimental measurements {[given in
  \eq{EXP1}]}. }
\label{fig:15tevtuning}
\end{center}
%
\begin{center} 
\includegraphics[width=0.45\textwidth, bb= 0 0 356 346]{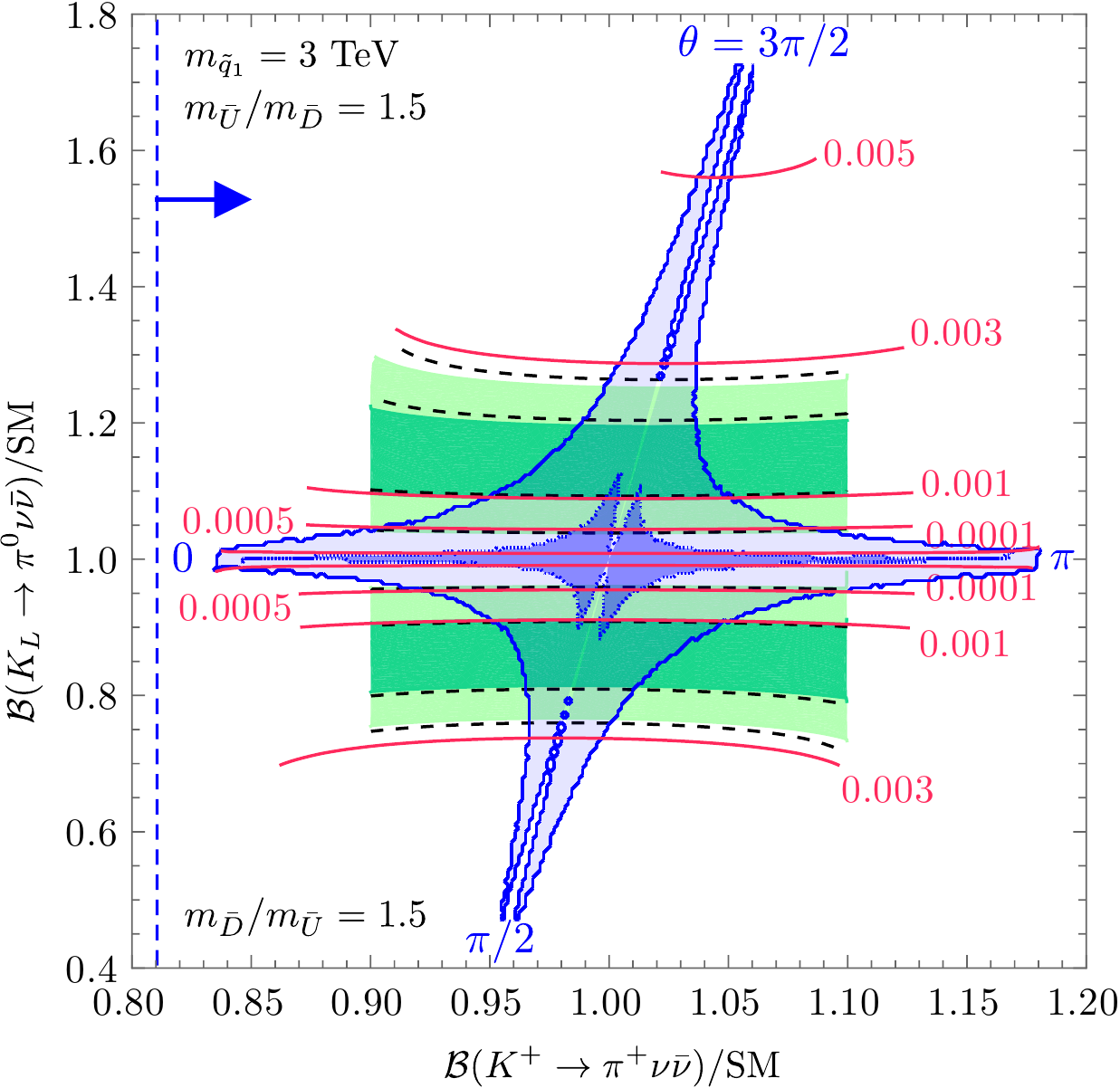}
\includegraphics[width=0.45\textwidth, bb= 0 0 356 346]{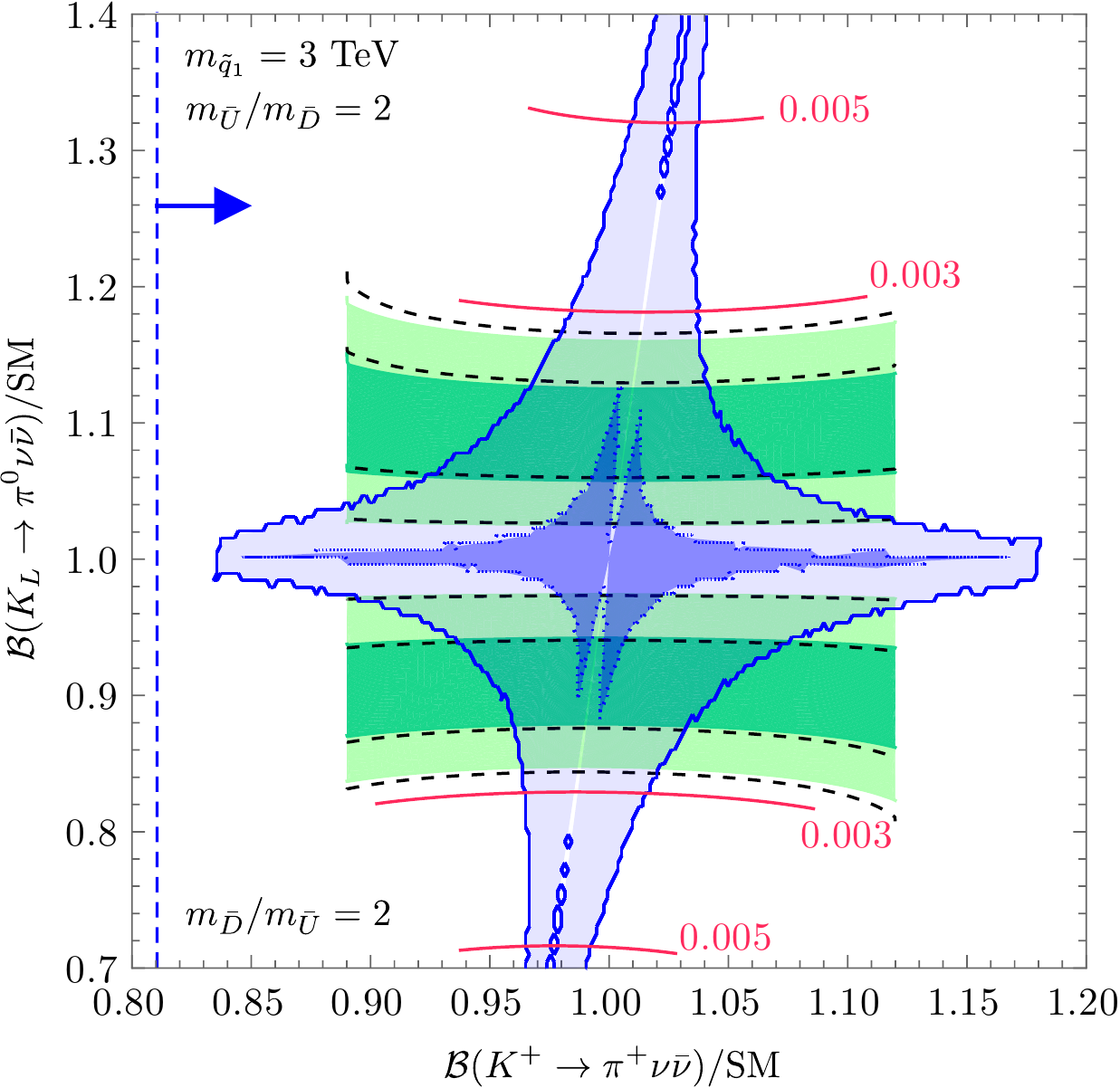}
\caption{
Same as Fig.~\ref{fig:15tevtuning} but for $m_{\tilde{q}_1} = 3\,\tev$,  
$m_{\bar{D}}/m_{\bar{U}}$\,(or $m_{\bar{U}}/m_{\bar{D}}) = 1.5$ (\emph{left} panel) 
and  $m_{\bar{D}}/m_{\bar{U}}$\,(or $m_{\bar{U}}/m_{\bar{D}}) = 2$ (\emph{right}).
}
\label{fig:3tevtuning}
\end{center}
\end{figure*}

Let us now investigate the degree of fine-tuning of the gluino mass
needed to suppress $\epsilon^{\rm SUSY}_K$.  In
Fig.~\ref{fig:15tevtuning}, the necessary amount of the fine-tuning
in the gluino mass with respect to the value for the exact cancellation
is shown, again in the $\mathcal{B}(K_L\to\pi^{0}
\nu\overline{\nu})$--$\mathcal{B}(K^{+}\to\pi^{+} \nu\overline{\nu})$
plane like Fig.~\ref{fig:M3MScontour}. In the light (dark) blue regions,
the amount of fine-tuning is milder than 1\,\% (10\,\%),
respectively while in the regions outside more sever fine-tuning is
required in order to satisfy constraints from $\epsilon_K$ (using the
inclusive $|V_{cb}|$ \cite{Bailey:2015tba}) and $ \Delta M_K$ at the
$2\,\sigma$ level. This means that the gluino mass can be shifted from
its value necessary for an exact cancellation in $\epsilon_K$ and
$\Delta M_K$ (given by the contours in Fig.~\ref{fig:M3MScontour}) by
1\,\% ($\sim20\,\gev$) and 10\,\% ($\sim200\,\gev$) 
without violating the constraints. Here we have scanned over
  all values of the CP violating phase $\theta$.
Alternatively, one can satisfy $\epsilon_K$ by tuning $\theta$
to values different from $\pm 90^\circ$ by at most $0.9^{\circ}$
  ($9^{\circ}$). In our plot we have discarded such a tuned CP phase,
  which explains the white dents in the blue regions around 
 $\theta=\pm 90^\circ$. 
 However, in the case of a tuned $\theta$ the correlation between the
 two $K\to \pi \nu\overline{\nu}$ branching ratios is stronger.  The red
contour show the SUSY contributions to $\epsilon^{\prime}_K /
\epsilon_K$ and the current $\epsilon_K^\prime/\epsilon_K$ discrepancy
is resolved at $1\,\sigma$\,($2\,\sigma$) within the dark (light) green
region.  The black dashed lines indicate the shifts of the boundaries of
the green regions when the gluino is taken to be $10\%$ heavier than
in Fig.~\ref{fig:M3MScontour}.  The lightest squark mass is fixed to
1.5\,\tev.  In the left (right) panel, we used $m_{\bar{D}}/m_{\bar{U}} =
1.1$ (2) with $m_{\bar{U}} = m_Q$ for $0 < \theta < \pi$, and
$m_{\bar{U}}/m_{\bar{D}} = 1.1$ (2) with $m_{\bar{D}} = m_Q$ for $\pi <
\theta < 2 \pi$.  The same results are depicted in
Fig.~\ref{fig:3tevtuning} but for a lightest squark mass of 3\,\tev, and
$m_{\bar{D}}/m_{\bar{U}} = 1.5$ (2) with $m_{\bar{U}} = m_Q$ is used for
$0 < \theta < \pi$, or $m_{\bar{U}}/m_{\bar{D}} = 1.5$ (2) with
$m_{\bar{D}} = m_Q$, in the left (right) panel.

Comparing Fig.~\ref{fig:15tevtuning} to Fig.~\ref{fig:3tevtuning} we can
see that if $m_{\bar{U}} / m_{\bar{D}} $ (or $m_{\bar{D}} / m_{\bar{U}}
$) differs more strongly from 1, $ | \mathcal{B}(K_L \to
\pi^0\nu\overline{\nu}) - \mathcal{B}^{\rm SM}(K_L \to
\pi^0\nu\overline{\nu}) |$ is predicted to be smaller in light of the
$\epsilon^{\prime}_K / \epsilon_K$ discrepancy. 
Figs.\,\ref{fig:15tevtuning} and \ref{fig:3tevtuning} also illustrate
 an important finding: There is
a strict correlation between $\mathcal{B}(K_L \to
\pi^0\nu\overline{\nu})$ and $m_{\bar{U}}/m_{\bar{D}}$: $\mbox{sgn}\,
(\mathcal{B}(K_L \to \pi^0\nu\overline{\nu})-\mathcal{B}^{\rm SM} (K_L
\to \pi^0\nu\overline{\nu}) ) = \mbox{sgn}\,(m_{\bar{U}}-m_{\bar{D}}) $.
This finding is easily understood by recalling that 
$\mbox{sgn}\,(m_{\bar{U}}-m_{\bar{D}})$ determines whether we must choose 
the CP phase $\theta$ between 0 and $\pi$ or instead between $\pi$ and
$2\pi$ to generate the desired positive  contribution to
$\epsilon_K^\prime$. Now the sign of the MSSM contribution to 
$\mathcal{B}(K_L \to \pi^0\nu\overline{\nu}) $ depends on the CP
phase in the same way, but there is no explicit dependence 
of $\mathcal{B}(K_L \to \pi^0\nu\overline{\nu}) $ on $m_{\bar{U},\bar{D}}$.
{The shape of the blue regions in Figs.\,\ref{fig:15tevtuning} and \ref{fig:3tevtuning} 
is a generic feature of NP models with FCNC transitions only among left-handed quarks 
and stems  from the constraint of $\epsilon_K$ on the new CP  phases \cite{Blanke:2009pq}.}

Numerically, we observed $\mathcal{B}(K_L \to
\pi^0\nu\overline{\nu})/\mathcal{B}^{\rm SM} (K_L \to
\pi^0\nu\overline{\nu})\lesssim 2\,(1.2)$ and $\mathcal{B}(K^+ \to
\pi^+\nu\overline{\nu})/\mathcal{B}^{\rm SM}(K^+ \to
\pi^+\nu\overline{\nu}) \lesssim 1.4\,(1.1)$ in light of
$\epsilon^{\prime}_K / \epsilon_K$ discrepancy, if all squark are
heavier than $1.5\,\tev$ and if a $1\%\,(10\%)$ fine-tuning is
permitted.  Here and hereafter, the quoted fine-tuning corresponds
  to the fine-tuning of the gluino mass or, alternatively, the tuning of
  the CP violating phase.  Similarly, $\mathcal{B}(K_L \to
\pi^0\nu\overline{\nu})/\mathcal{B}^{\rm SM} (K_L \to
\pi^0\nu\overline{\nu})\lesssim 1.1$ and $\mathcal{B}(K^+ \to
\pi^+\nu\overline{\nu})/\mathcal{B}^{\rm SM}(K^+ \to
\pi^+\nu\overline{\nu}) \lesssim 1.02$ are predicted, if all squark
masses are above $3\,\tev$ with a $10\%$ fine-tuning.

  Note that if $m_{\bar{U}}/m_{\bar{D}} $ is close to $1$, the Trojan
  penguin contribution from the SUSY QCD box diagrams are
  suppressed and the gluino contribution to the chromomagnetic operator
  entering $\epsilon^{\prime}_K / \epsilon_K$ becomes dominant: for
  $m_{\bar{U}}/m_{\bar{D}} $ = 1.05 (1.02), 25 \,\% (50\,\%) of the SUSY
  contribution comes from the chromomagnetic operator for
  $m_{\tilde{q}_1} = 1.5\,\tev$ and larger values of $| \mathcal{B}(K_L
  \to \pi^0\nu\overline{\nu}) - \mathcal{B}^{\rm SM}(K_L \to
  \pi^0\nu\overline{\nu}) |$ are predicted. However, it is shown that
  such a case always requires fine-tuning at the $1\%$ level.

\section{Discussion and Conclusions}
\label{conclusions}

In this article, we have studied the correlations between $\epsilon_K$,
$\epsilon_K^\prime$, $K_L \to \pi^0\nu\overline{\nu}$ and $K^+ \to
\pi^+\nu\overline{\nu}$ in detail within the MSSM. In order to
accommodate the $\epsilon_K^\prime/\epsilon_K$ anomaly, we generate
isospin violation by a mass splitting between right-handed up and
down-squark and flavor as well as CP violating by off-diagonal elements
in the left-handed bilinear squark mass terms.

We find strong correlations between these observables depending (to a
very good approximation) only on
$m_Q,~|\Delta_{Q,12}|,~\theta,~M_3,~M_2,~m_{\bar{U}}/m_{\bar{D}},~m_L$. In
particular, we find the following prediction: $\mbox{sgn}\,
(\mathcal{B}(K_L \to \pi^0\nu\overline{\nu})-\mathcal{B}^{\rm SM} (K_L
\to \pi^0\nu\overline{\nu}) ) = \mbox{sgn}\,(m_{\bar{U}}-m_{\bar{D}})
$. This is in contrast to generic $Z^\prime$ models where couplings to
leptons are in general free parameters, decoupling
$\epsilon_K^\prime/\epsilon_K$ from $K_L \to \pi^0\nu\overline{\nu}$ and
$K^+ \to \pi^+\nu\overline{\nu}$.

We show that $ \mathcal{B}(K_L \to \pi^{0} \nu \overline{\nu} ) $ is expected to be shifted with respect to the SM value by 5\%--10\% within the typical parameter region of our scenario.
 Even a larger shift is possible if one allows for fine tuning: $\mathcal{B}(K_L \to \pi^0\nu\overline{\nu})/\mathcal{B}^{\rm SM} (K_L \to \pi^0\nu\overline{\nu})\lesssim 2\,(1.2)$ and
  $\mathcal{B}(K^+ \to \pi^+\nu\overline{\nu})/\mathcal{B}^{\rm SM}(K^+
  \to \pi^+\nu\overline{\nu}) \lesssim 1.4\,(1.1)$ 
for a fine-tuning at the $1\%\,(10\%)$ level. 

It is also clearly shown that our scenario can be distinguished from
those with dominant $Z$-penguins.  In the latter scenarios, the
$Z$-penguin contributions to $\epsilon^{\prime}_K $ is proportional to
$(\textrm{Im}{\Delta_L} + 3.3 \,\textrm{Im} { \Delta_R})$ and
$\mathcal{B}(K_L \to \pi^{0} \nu \overline{\nu} ) - \mathcal{B}^{\rm
  SM}(K_L \to \pi^{0} \nu \overline{\nu} )$ is proportional to $ -
(\textrm{Im}{\Delta_L} + \textrm{Im} { \Delta_R})$. Therefore, a
suppression of the branching ratio of $K_L \to \pi^{0} \nu
\overline{\nu}$ (numerically $\mathcal{B}(K_L \to \pi^0 \nu
\overline{\nu})/\mathcal{B}^{\rm SM}(K_L \to \pi^0 \nu \overline{\nu})
\lesssim 0.7$ \cite{Endo:2016tnu}) is in general predicted if there is
no cancellation between $\textrm{Im}{\Delta_L} $ and
$\textrm{Im}{\Delta_R} $ \cite{Buras:2015yca}. Here, $\Delta_{L(R)} $
denotes the effective coupling of $\bar{s}\gamma_{\mu} P_{L(R)} d
Z^{\mu}$ originating from NP interactions. This means that an accurate
measurement of $K_L \to \pi^0\nu\overline{\nu}$ would be able to
distinguish these scenarios.

For our analysis, we assume GUT relations among the gauginos. Relaxing
this assumption allows for larger, but less correlated, effects in $K_L
\to \pi^0\nu\overline{\nu}$ and $K^+ \to \pi^+\nu\overline{\nu}$. Such
an analysis together with a presentation of the complete analytic
expressions for $\epsilon_K^\prime/\epsilon_K$, $K_L \to
\pi^0\nu\overline{\nu}$ and $K^+ \to \pi^+\nu\overline{\nu}$ will be
presented in a forthcoming article.

Finally, we discuss the Higgs boson mass within the MSSM.  In our
  phenomenological analysis ($\tan \beta = 10 $ and $A_{ij} =0$), the
  Higgs boson mass of 125$\,\gev$ can be achieved for stop masses around
  5\,\tev \,\cite{Hahn:2013ria}.  
  To accommodate for the measured Higgs mass with lighter stops we have checked that one can choose large diagonal trilinear $A_{ii}$ terms
 (defined with the Yukawa
  couplings factored out) without relevant effect on the studied
  observables.  In particular, diagonal $A_{ii}$ terms neither generate
  sizable $Z$-penguins nor effects in nucleon EDM.  Furthermore,
  promoting the MSSM to the NMSSM or adding additional $D$-term
  contributions to the Higgs boson mass would leave our analysis
  unchanged. (The patterns of flavor observables in the MSSM and NMSSM
  are essentially identical.)  Therefore, one can account for the
  measured value of 125\,GeV for the light Higgs boson mass within our
  setup.

{\it Acknowledgments} --- {\small A.~C. is supported by an Ambizione Grant
  (No.~PZ00P2\_154834) of the Swiss National Science Foundation. G.~D.
  was supported in part by MIUR under Project No.~2015P5SBHT and by the INFN
  research initiative ENP. The work of U.~N. is supported by the German Bundesministerium f\"ur Bildung und Forschung under
Grant No.~05H15VKKB1.  }

\end{document}